\documentstyle[epsf]{l-aa}
\def\sign{\mathop{\rm sign}\nolimits}

\def\arcsinh{\mathop{\rm arcsinh}\nolimits}
\def\arccosh{\mathop{\rm arccosh}\nolimits}
\def\arctanh{\mathop{\rm arctanh}\nolimits}

\newcommand{\etal}[1]{et al.}
\newcommand{\firstauthor}[2]{#2 #1}
\newcommand{\otherauthor}[2]{#2 #1}
\newcommand{\article}[9]%
{\bibitem[#1]{#2}#3, #9, #5 \underline{#6}, #8}
\newcommand{\novolume}[8]%
{\bibitem[#1]{#2}#3, #8, #5 \underline{#6}, #7}
\def\edtext{eds.}
\newcommand{\proceedings}[9]%
{\bibitem[#1]{#2}#3, #9, #4.\, In: #6 (\edtext) #5. #8, #7}

\newcommand{\sproceedings}[9]%
{\bibitem[#1]{#2}#3, #9, #4.\, In: #6 (\edtext) #5. #8, #7}
\def\preptext{in preparation}
\newcommand{\prep}[4]%
{\bibitem[#1]{#2}#3, #4, \preptext}
\def\subtext{submitted}
\newcommand{\submitted}[6]%
{\bibitem[#1]{#2}#3, #6, #5 (\subtext)}
\def\accepttext{in press}
\newcommand{\inpress}[6]%
{\bibitem[#1]{#2}#3, #6, #5 (\accepttext)}
\newcommand{\book}[7]%
{\bibitem[#1]{#2}#3, #7, #4.\, #6, #5}
\newcommand{\doctor}[6]%
{\bibitem[#1]{#2}#3, #6, doctoral thesis, #5}
\newcommand{\D}[1]{{\rm d}#1}                       
\begin{document}
\thesaurus{
           02        
          (12.03.4;  
           03.13.4   
           12.04.3;  
           12.03.2;  
           12.07.1)  
}
\title{
A general and practical method for calculating cosmological distances
} 
\author{
Rainer Kayser\thanks{\tt rkayser@hs.uni-hamburg.de}\inst{1}, 
Phillip Helbig\thanks{\tt phelbig@hs.uni-hamburg.de}\inst{1} \and  
Thomas Schramm\thanks{\tt schramm@tu-harburg.d400.de}\inst{1,2}
}
\offprints{
P.~Helbig
}
\institute{
           Hamburger Sternwarte,
           Gojenbergsweg 112,
           D-21029 Hamburg-Bergedorf, Germany \and
           Rechenzentrum, Technische Universit\"{a}t Hamburg-Harburg,
           Denickestra\ss e 17, D-21071  Hamburg-Harburg, Germany
}
\date{
accepted
}
\maketitle
\markboth{Kayser et~al.: Cosmological distances}{}
\begin{abstract}
The calculation of distances is of fundamental importance in
extragalactic astronomy and cosmology. However, no practical
implementation for the general case has previously been available.  We
derive a second-order differential equation for the angular size
distance valid not only in all {\em homogeneous\/} Friedmann-Lema\^\i
tre cosmological models, parametrised by $\lambda_{0}$ and $\Omega_{0}$,
but also in {\em inhomogeneous\/} `on-average' Friedmann-Lema\^\i tre
models, where the inhomogeneity is given by the (in the general case
redshift-dependent) parameter~$\eta$.  Since most other cosmological
distances can be obtained trivially from the angular size distance, and
since the differential equation can be efficiently solved numerically,
this offers for the first time a practical method for calculating
distances in a large class of cosmological models.  We also briefly
discuss our numerical implementation, which is publicly available. 
\keywords{cosmology: theory -- methods: numerical -- 
          cosmology: distance scale -- cosmology: miscellaneous -- 
          gravitational lensing
}
\end{abstract}

\section{Introduction}\label{intro}

The determination of distances is one of the most important problems in
extragalactic astronomy and cosmology.  Distances between two objects X
and Y depend on their redshifts~$z_{x}$ and $z_{y}$, the Hubble
constant~$H_{0}$, the cosmological constant $\lambda_{0}$, the density
parameter $\Omega_{0}$ and the inhomogeneity
parameter~$\eta$.\footnote{When discussing the distance between {\em
two\/} objects, one can always make a coordinate transformation such
that the contribution from the $\theta$ and $\phi$ terms in
Eq.~(\ref{rwm-g}) vanish.  Then one simply needs the redshifts and
cosmological parameters in order to determine the distance between them.
When discussing the distances between several objects, for example QSOs
with $\alpha$, $\delta$ and $z$ as coordinates, this is no longer
possible.  In many cases, however, suitable geometrical approximations
can be made so that the most complicated part of the problem is
essentially a determination of a distance between two objects. This
point is further discussed in Sect.~\ref{numerics}.} Usually, smaller
distances are determined by the traditional `distance ladder' technique
and larger distances are calculated from the redshift, assuming some
cosmological model.  Since the redshift is for most purposes exactly
measurable, knowledge of or assumptions about two of the
factors~(a)~Hubble constant, (b)~other cosmological parameters and
(c)~`astronomical distance' (i.e.~ultimately tied in to the local
distance scale) determines the third.  In this paper we discuss
distances given the Hubble constant $H_{0}$,  the redshifts $z_{x}$ and
$z_{y}$ and the cosmological parameters $\lambda_{0}$, $\Omega_{0}$ and
$\eta$. Traditionally, a simple cosmological model is often assumed for
ease of calculation, although the distances thus obtained, and results
which depend on them, might be false if the assumed cosmological model
does not appropriately describe our universe. A general method allows
one to look at cosmological models whether or not they are
easy-to-calculate special cases and offers the possibility of
determining cosmological distances which are important for other
astrophysical topics once the correct cosmological model is known. 

We stress the fact that the inhomogeneity can be as important as the
other cosmological parameters, both in the field of more traditional
cosmology and in the case of gravitational lensing, where, e.g.~in the
case of the time delay between the different images of a multiply imaged
source, the inhomogeneity cannot be neglected in a thorough analysis
(Kayser \& Refsdal \cite{RKayserSRefsdal83a}). For an example involving
a more traditional cosmological test, Perlmutter et~al.\
(\cite{SPerlmutter95a}) (see also Goobar \& Perlmutter 
(\cite{AGoobarSPerlmutter95a})) discuss using supernovae with $z \approx
0.25$--$0.5$ to determine $q_{0}$; for $z$ near the top of this range or
larger, the uncertainty due to our ignorance of $\eta$ is comparable
with the other uncertainties of the method. 

The plan of this paper is as follows.  In Sect.~\ref{cosmology} the
basics of Friedmann-Lema\^\i tre cosmology are briefly discussed; this
also serves to define our terms, which is important since various
conflicting notational schemes are in use.  (For a more thorough
discussion using a similar notation see, e.g., Feige
(\cite{BFeige92a}).) Section~\ref{distances} defines the various
distances used in cosmology. In Sect.~\ref{rainer} our new differential
equation is derived.  Similar efforts in the literature are briefly
discussed.  Section~\ref{numerics} briefly describes our numerical
implementation and gives the details on how to obtain the source code
for use as a `black box' (which however can be opened) for use in
cosmology and extragalactic astronomy.  The symmetry properties of the
angular size distance, analytic solutions and methods of calculating the
volume element are addressed in three appendices.

\section{Basic theory}\label{cosmology}

Considering for the moment {\em homogeneous\/} Friedmann-Lema\^\i tre
cosmological models, we can write the familiar Robertson-Walker line
element: 
\begin{eqnarray}\label{rwm-g}
ds^{2} & = & c^{2} {\rm d}t^{2} - R^{2}(t) \quad \times \nonumber \\
       &   & 
       \left(\frac{{\rm d}\sigma^{2}}{\left(1-k\sigma^{2}\right)}
         + \sigma^{2}{\rm d}\theta^{2} + 
           \sigma^{2}\sin^{2}\theta {\rm d}\phi^{2} \right),
\end{eqnarray}
where the symbols are defined as follows (with the corresponding units):
\begin{description}
\item[$s$] 4-dimensional interval\hfill [length]
\item[$c$] speed of light\hfill [velocity]
\item[$t$] time\hfill [time]
\item[$R$] scale factor\hfill [length]
\item[$\sigma$] radial coordinate\hfill [dimensionless]
\item[$k$] curvature constant\hfill [dimensionless]
\item[$\theta$] angular coordinate\hfill [dimensionless]
\item[$\phi$] angular coordinate\hfill [dimensionless]
\end{description}
The dynamics of the universe is given by the Friedmann equations
\begin{equation}\label{flg1-g}
\dot R^{2}(t) = \frac{8\pi G\rho(t)R^{2}(t)}{3} +
\frac{\Lambda R^{2}(t)}{3} - kc^{2}
\end{equation}
and
\begin{equation}\label{flg2-g}
\frac{\ddot R(t)}{R(t)} = -\frac{4\pi G\rho(t)}{3}+\frac{\Lambda}{3},
\end{equation}
where dots denote derivatives with respect to $t$, $G$ is the
gravitational constant,~$\rho(t)$ the matter density (this paper assumes
negligible pressure), $\Lambda$  the cosmological constant and the sign
of $k$ determines the curvature of the 3-dimensional space. 

Introducing the usual parameters 
\newlength{\lambdalength}
\settowidth{\lambdalength}{(normalised cosmological constant)}
\newlength{\hlen}
\settowidth{\hlen}{$\displaystyle\frac{\dot R}{R}$}
\newlength{\olen}
\settowidth{\olen}{$\displaystyle\frac{8\pi G\rho}{3H^{2}}$}
\newlength{\llen}
\settowidth{\llen}{$\displaystyle\frac{\Lambda}{3H^{2}}$}
\begin{eqnarray}
\label{HOL-g}
H        & = & \frac{\dot R}{R}
\settowidth{\lambdalength}{(normalised cosmological constant)}
\addtolength{\lambdalength}{\olen}
\addtolength{\lambdalength}{\llen}
\makebox[\lambdalength][r]{(Hubble parameter)}
\nonumber
\\ 
\Omega   & = & \frac{8\pi G\rho}{3H^{2}}
\settowidth{\lambdalength}{(normalised cosmological constant)}
\addtolength{\lambdalength}{\hlen}
\addtolength{\lambdalength}{\llen}
\makebox[\lambdalength][r]{(density parameter)}
\\
\lambda  & = & \frac{\Lambda}{3H^{2}}
\settowidth{\lambdalength}{(normalised cosmological constant)}
\addtolength{\lambdalength}{\hlen}
\addtolength{\lambdalength}{\olen}
\makebox[\lambdalength][r]{(normalised cosmological constant)}
\nonumber
\end{eqnarray}
($\Omega$ and $\lambda$ are dimensionless and $H$ has the dimension
$t^{-1}$) we can use Eq.~(\ref{flg1-g}) to calculate 
\begin{equation}\label{kc-g}
kc^{2} = R^{2}H^{2}\left(\Omega + \lambda - 1\right),
\end{equation}
so that
\begin{equation}
\label{k-g}
k = \sign\left(\Omega + \lambda -1\right).
\end{equation}
Since $R>0$ we can write
\begin{equation}\label{rnull-g}
R = \frac{c}{H}\frac{1}{\sqrt{|\Omega + \lambda -1|}};
\end{equation}
this is the radius of curvature of the 3-dimensional space at time~$t$.
For $k=0$ it is convenient to {\em define\/} the scale factor $R$ to be
$c/H$.  In the following the index~$0$ will be used to denote the
present value of a given quantity, fixed, as usual, at the time~$t_{0}$ of 
observation.\footnote{Note that this paper is concerned with the 
calculation of distances from redshift.  We are not concerned with a 
change in redshift with $t_0$.}
The explicit dependence on $t$ will be dropped for brevity.  Taking matter 
conservation into account 
and using the present-day values, we have
\begin{equation}
\label{conserve-g}
\rho R^{3} = \rho_{0}R_{0}^{3}
\end{equation}
and so from Eqs.~(\ref{flg1-g}), (\ref{HOL-g}), (\ref{kc-g}) and 
(\ref{conserve-g}) 
follows
\begin{equation}\label{flg-a}
\dot R^{2} = H_{0}^{2}R_{0}^{2}
             \left(
             \frac{\Omega_{0}R_{0}}{R} +
             \frac{ \lambda_{0}R^{2}}{R_{0}^{2}} -
             (\Omega_{0} + \lambda_{0} - 1)  
             \right).
\end{equation}
Since below we want to discuss distances as functions of the
cosmological redshift~$z$, by making use of the facts that 
\begin{equation}
\label{lemaitre-g} z = \frac{R_{0}}{R} - 1
\end{equation}
and that $R_{0}$ is fixed,
we can use Eq.~(\ref{flg-a}) to get
\begin{equation}\label{zpunkt-g}
\D z = \frac{\D z}{\D R}\dot R \D t = - H_{0}(1 + z)\sqrt{Q(z)}\,\D t ,
\end{equation}
where
\begin{equation}\label{q-g}
Q(z) = \Omega_{0}(1 + z)^{3} - 
       (\Omega_{0} + \lambda_{0} -1)(1 + z)^{2} +
       \lambda_{0} .
\end{equation}

{\bf Note:} {\em Throughout this paper, the~$\sqrt{\qquad}$~sign should
be taken to signify the {\em positive} solution, except that
$\sign\sqrt{Q(z)}=\sign(\dot R)$ always.}

\section{Distance measures}\label{distances}

\subsection{Distances defined by measurement}

In a static Euclidean space, one can define a variety of distances
according to the method of measurement, which are all equivalent.

\subsubsection{Angular size distance}

Let us consider at position $y$ two light rays intersecting at $x$ with
angle $\theta$. If $l$ is the distance between these light rays, it is
meaningful to define the angular size distance~$D_{xy}$ as 
\begin{equation}\label{d-g}
D_{xy} = \frac{l}{\theta}, 
\end{equation}
since an object of projected length~$l$ at position $y$ will subtend an
angle~$\theta=l/D_{xy}$ (for small $\theta$) at distance~$D_{xy}$.

\subsubsection{Proper motion distance}

The proper motion distance is similar to the angular size distance,
except that $l$ is given by $vt$, where $v$ is the tangential velocity
of an object and $t$ the time during which the proper motion is
measured.

\subsubsection{Parallax distance}

Parallax distance is similar to the proper motion distance, except that
the angle $\pi$ is at $y$ instead of $x$, so that we have 
\begin{equation}
D^{\pi}\!\!_{xy} = \frac{l}{\pi}.
\end{equation}
In the canonical case, $l=1~{\rm AU}$.

\subsubsection{Luminosity distance}

Since the apparent luminosity~$L$ of an object at distance $D$ is
proportional to $1/D^{2}$, one can define the luminosity distance as 
\begin{equation}
D^{\rm L} = D^{\rm L}_{0}\sqrt{\frac{L_{0}}{L}},
\end{equation}
where $L_{0}$ is the luminosity at some fiducial distance $D^{\rm
L}_{0}$.

\subsubsection{Proper distance}

By proper distance~$D^{\rm P}$ we mean the distance measured with a
rigid ruler.

\subsubsection{Distance by light travel time}

Finally, from the time required for light to traverse a certain
distance, one can define a distance $D^{\rm c}$ by 
\begin{equation}
D^{\rm c} = ct
\end{equation}
where $t$ is the so-called look-back time.

\subsection{Cosmological distances}

\subsubsection{General considerations}

In a static Euclidean space, which was used above when defining the
distances through a measurement description, these distance measures are
of course equivalent.  In the general case in cosmology, where the
3-dimensional space need not be flat ($k=0$) but can be either
positively ($k=+1$) or negatively ($k=-1$) curved, and where the
3-dimensional space is scaled by $R(t)$, not only do the distances
defined above differ, but also (in the general case) $D_{xy}\neq
D_{yx}$.  The definitions are still applicable, but different
definitions will result in different distances. 

In reality, of course, the universe is neither perfectly homogeneous nor
perfectly isotropic, as one assumes when deriving Eq.~(\ref{rwm-g}).
However, as far as the usefulness of the Friedmann equations in
determining the global dynamics is concerned, this appears to be a good
approximation.  (See, for example, Longair (\cite{MLongair93a}) and
references therein for an interesting discussion.)  The approximation is
certainly too crude when using the cosmological model to determine
distances as a function of redshift, since the angles involved in such
cases can have a scale comparable to that of the inhomogeneities.  In
this paper, we assume that these inhomogeneities can be sufficiently
accurately described by the parameter~$\eta$, which gives the fraction
of homogeneously distributed matter.  The rest ($1 - \eta$) of the
matter is distributed clumpily, where the scale of the clumpiness is
{\em by definition\/} of the same order of magnitude as the angles
involved. 

For example, a halo of compact MACHO type objects around a galaxy in a
distant cluster would be counted among the homogeneously distributed
matter if one were concerned with the angular size distance to
background galaxies further away, but would be considered clumped on
scales such as those important when considering microlensing by the
compact objects themselves.  Since we don't know exactly how dark matter
is distributed, different $\eta$ values can be examined to get an idea
as to how this uncertainty affects whatever it is one is interested in. 
If one has no selection effects, then, due to flux conservation, the
`average' distance cannot change (Weinberg \cite{SWeinberg76a}); $\eta$
introduces an additional uncertainty when interpreting observations.  It
is generally not possible to estimate this scatter by comparing the
cases $\eta=0$ and $\eta=1$, since, depending on the cosmological
parameters and the cosmological mass distribution, not all combinations
are self-consistent.  For instance, if one looks at scales where
galaxies are compact objects, and the fraction of $\Omega_{0}$ due to
the galaxies is $x$, then $\eta$ {\em must\/} be $\leq (1-x)$. 

We further assume that light rays from the object whose distance is to
be determined propagate sufficiently far from all clumps.  (See
Schneider et al.\ (\cite{PSchneiderEF92a}) -- hereafter SEF -- for a
more thorough discussion of this point.)  Compared to the perfectly
homogeneous and isotropic case, the introduction of the $\eta$ parameter
will influence the angular size and luminosity distances (as well as the
proper motion and parallax distances) since these depend on angles
between light rays which are influenced by the amount of matter in the
beam, but not the proper distance and only negligibly the light travel
time.  The last two distances are discussed briefly in
Sect.~\ref{relation} and in App.~\ref{eta1} and \ref{time}.  Since there
is a simple relation between the angular size distance and the
luminosity distance (Sect.~\ref{relation}) which also holds for the
inhomogeneous case (see App.~\ref{symmetry}), for the general case it
suffices to discuss the angular size distance, which we do in
Sect.~\ref{rainer}.

\subsubsection{Relationships between different distances}\label{relation}

Without derivation\footnote{See, e.g., Feige (\cite{BFeige92a}) Berry
(\cite{MBerry86a}) or Bondi (\cite{HBondi61a})  for a more general
discussion.  What we present in the rest of this section is not new, but
is important in order to clarify the notation.  The results are obvious
from the definitions introduced above.} we now discuss some important
distance measures, denoting the redshifts of the objects with the
indices $x$ and $y$.  Due to symmetry considerations (see
App.~\ref{symmetry}) 
\begin{equation}\label{angrel-g} 
D_{yx} = D_{xy}\left(\frac{1 + z_{y}}{1 + z_{x}}\right), 
\end{equation} 
where the term in parentheses takes account of, by way of
Eq.~(\ref{lemaitre-g}), the expansion of the universe.  It is convenient,
in keeping with the meaning of angular size distance, to think of the
expansion of the universe changing the angle $\theta$ in Eq.~(\ref{d-g})
and not $l$, if one identifies $l$ as the (projected) size of an object.
The angle is defined at the time when the light rays intersect the
plane of the observer.  Thus $D_{xy}$ with the observer at $x=0$ defines
what one normally thinks of as an angular size distance.  On the other
hand, $D_{xy}$ and $D_{yx}$ with $x$ in general $\neq0$ can be important
in, for example, gravitational lensing.\footnote{Although not useful in
cosmology or extragalactic astronomy, for completeness we mention the
fact that the proper motion distance is equivalent to
$D_{yx}$ and the parallax distance is equivalent to 
$D_{yx}/\sqrt{1 - k\sigma^{2}}$.}

Although the angle between the rays (at the source) at the time of
reception of the light is important for the luminosity distance, this
distance is {\em not\/} simply $D_{yx}$, since in the cosmological case
the observed flux is obtained by multiplying the `non-redshifted flux'
by the factor $(1+z_{x})^{2}/(1+z_{y})^{2}$. One factor of
$(1+z_{x})/(1+z_{y})$ occurs because a given wavelength is increased by
$(1+z_{y})/(1+z_{x})$, which reduces the flux correspondingly; an
additional factor of $(1+z_{x})/(1+z_{y})$ occurs because the arrival
rate of photons is also decreased.  Therefore, since $D^{\rm L}$ is
inversely proportional to the square root of the (observed,
`redshifted') flux the luminosity distance is 
\begin{equation}\label{lum-g}
D^{\rm L}_{xy} = D_{yx}\left(\frac{1+z_{y}}{1+z_{x}}\right).
\end{equation}
From this and Eq.~(\ref{angrel-g}) follows the relation
\begin{equation}\label{lumang-g}
D^{\rm L}_{xy} = D_{xy}\left(\frac{1+z_{y}}{1+z_{x}}\right)^{2}.
\end{equation} 
This means that the surface brightness of a `standard candle'
is~\mbox{$\sim(1+z)^{-4}$}, a result independent of the cosmological
model parameters, including $\eta$.\footnote{Thus, a `surface brightness
test' can in principle show that cosmological redshifts are due to the
expansion of the universe and not to some other cause.  See, e.g.,
Sect.~6 in Sandage (\cite{ASandage95a}).}  (This result also holds for
the inhomogeneous case, since Eq.~(\ref{angrel-g}) still holds (see
App.~\ref{symmetry}) and the additional factor due to the expansion of
the universe (given by the term in parentheses in Eq.~(\ref{lum-g})) is
of course present in the inhomogeneous case as well.) 

Of course, this applies only to the {\em bolometric\/} luminosity.
Observing in a finite band introduces two corrections.  The so-called
$K$-correction as it is usually defined today (see, e.g., Coleman et
al.\ (\cite{CColemanWW80a}) or, for an interesting and thorough
discussion, Sandage (\cite{ASandage95a})) takes account of these, both
of which come from the fact that the observed wavelength interval is
redshifted compared to the corresponding interval on emission.  This
means that, first, for a flat spectrum, less radiation is observed,
because the bandwidth at the observer is $(1+z)$ times larger than at
the source. Second, the spectrum need not be flat, in which case
additional corrections based on the shape of the spectrum have to be
included.\footnote{Since the observed objects generally evolve with
time, and redshifted objects are necessarily observed as they were when
the radiation was emitted, some authors include an evolutionary term in
the $K$-correction.  Still other authors prefer to absorb one or more of
these terms into the definition of the luminosity distance.  Our
luminosity distance is a bolometric distance based on the geometry and
includes the unavoidable dimming due to the redshift.  Our
$K$-correction takes account of both effects of a finite bandwidth.
Evolutionary effects are considered separately from distances.} Thus, 
\begin{equation}
m =  M + 5\log \left(\frac{D^{\rm L} [{\rm pc}]}{10\:\rm{pc}}\right) + K
\end{equation}
where $m$ is the apparent magnitude, $M$ the absolute magnitude, $D^{\rm
L}$ is the luminosity distance and $K$ is the $K$-correction as defined
in Coleman et al.\ (\cite{CColemanWW80a}). 
Perhaps more convenient is 
\begin{equation}
m =  M + 5\log D^{\rm L} + K + N
\end{equation}
where $N$ is a normalisation term: $N = -5$ for $D^{\rm L}$ in units of
1~pc, $N = 25$ for $D^{\rm L}$ in units of 1~Mpc and $N = x - 5\log h$
for $D^{\rm L}$ in units of the Hubble length\footnote{For example, as
given by our numerical implementation; see Sect.~\ref{numerics}}
$c/H_{0}$, where 
\begin{displaymath}
x = 5\log\left(\frac{\mbox{Hubble length}}{\mbox{1 pc}}\right)-5\approx 42.384
\end{displaymath}
and $h$ is the Hubble constant in units of $100\thinspace \rm km/s/Mpc$.
In practice one has to add terms to correct for various sources of
extinction and consider the fact that $M$ is the absolute magnitude of
the object when the light was emitted, which of course could be
different from the present $M$ of similar objects at negligible
redshift. 

The light travel time (or lookback time) $t_{xy}=t_{x}-t_{y}$ between
$z_x$ and $z_y$ (where $t_x=t(z_x)>t_y =t(z_y)$) is given by the
integration of the reciprocal of Eq.~(\ref{zpunkt-g}): 
\begin{equation}\label{t-g}
t_{xy}=\int\limits_{z_y}^{z_x}\left(\frac{\D{z}}{\D{t}}\right)^{-1}\D{z}
      =\frac{1}{H_0}\int\limits_{z_x}^{z_y}\frac{\D{z}}{(1+z)\sqrt{Q(z)}},
\end{equation}
where the minus sign from Eq.~(\ref{zpunkt-g}) is equivalent to the
swapped limits of integration on the right-hand side so that the
integral gives $t_{x}-t_{y}$ instead of $t_{y}-t_{x}$, making the light
travel time increase (for $\dot R > 0$) with $z$; thus $D^{\rm c}_{xy} =
ct_{xy}$. 

Since the proper distance would be the same as $D^{\rm c}$ {\em were
there no expansion}, the former can be calculated by multiplying the
integrand in Eq.~(\ref{t-g}) by $c(1 + z)$.  Thus 
\begin{equation}\label{dp-g}
D^{\rm P}_{xy} = \frac{c}{H_{0}}\int\limits_{z_{x}}^{z_{y}}
\frac{{\rm d}z}{\sqrt{Q(z)}}.
\end{equation}
This gives the proper distance at the present time.  Since $D^{\rm P}$
scales linearly with the expansion of the universe, the proper distance
at some other time can be obtained by dividing Eq.~(\ref{dp-g}) with
$(1+ z_{i})$, where $z_{i}$ is the redshift at the corresponding time.
For homogeneous ($\eta=1$) cosmological models,\footnote{This includes
{\em empty\/} models ($\Omega_{0} = 0$); although $\eta$ has no meaning
here, the same arguments apply.} the propagation of light rays is
determined by the global geometry, so that there is a simple relation
between $D^{\rm P}$ and $D$ and, thus, $D^{\rm L}$.  This is discussed
in Sect.~\ref{eta1}.  Although not `directly' observable, the proper
distance is nevertheless important in cosmological theory, since it is
the basic distance of general relativity.  Although not useful as a
distance, the light travel time is of course important when considering
evolutionary effects. 

For inhomogeneous models, where this relation between global geometry
and local light propagation does not exist, another approach must be
used, which takes account of both the expansion of the universe as well
as the local propagation of light, when calculating angle-defined
distances such as the angular size distance.

\section{The general differential equation for the angular size distance}
\label{rainer}

In a series of papers Zeldovich (\cite{YZeldovich64a}), Dashevskii and
Zeldovich (\cite{VDashevskiiYZeldovich65a}) and Dashevskii and Slysh
(\cite{VDashevskiiVSlysh66a}) developed a general differential equation
for the distance between two light rays on the boundary of a small light
cone propagating far away from all clumps of matter in an inhomogeneous
universe: 
\begin{equation}\label{zds-g}
\ddot{l} = -4\pi G\eta\rho \,l + {\dot{R}\over R}\,\dot{l} 
\end{equation}
where $\eta$ and $\rho$ are functions of the time $t$ ({\em not\/} the
lookback time of Eq.~\ref{t-g}). The first term can be interpreted as
Ricci focusing due to the matter inside the light cone, and the second
term is due to the expansion of space during the light propagation. We
now have to transform this {\em time\/} dependent differential equation
into a {\em redshift\/} dependent differential equation.  From
Eq.~(\ref{zpunkt-g}) we obtain\footnote{This transformation causes
problems if the integration interval contains a point where $\dot R=0$
and thus $\sqrt{Q}$ changes sign.  In this case the integration interval
$(t_x,t_y)$ has to be transformed into {\em two\/} integration
intervals, namely $(z_x,z_{\rm max})$ and $(z_{\rm max},z_y)$, where
$z_{\rm max}$ is the redshift at $\dot R=0$, with the boundary
conditions for the second integration interval chosen appropriately.} 
\begin{equation}\label{RKeqA}
\D t = - \left(H_0 (1+z) \sqrt{Q}\right)^{-1}\,\D z,
\end{equation}
and thus
\begin{equation}\label{RKeqB}
{\D l\over\D t} = - H_0 (1+z)\sqrt{Q}\,{\D l\over\D z}
\end{equation}
and
\begin{eqnarray}
{\D^2 l\over\D t^2} & = & 
 H_0^2 (1+z) \sqrt{Q} {\D\over\D z}\left((1+z)\sqrt{Q}\,
  {\D l\over\D z}\right) \\
 & = & H_0^2 \left(\left((1+z)Q + (1+z)^2 {1\over 2}\,{\D Q\over\D z}\right)\,
     {\D l\over\D z}\right.  \nonumber\\ 
 & & \left. +\quad (1+z)^2 Q \, {\D^2 l\over\D z^2}\right)\label{RKeqC}\quad .
\end{eqnarray}
Furthermore, since $R = R_0/\!(1+z)$ 
(Eq.~(\ref{lemaitre-g})), we obtain, using Eq.~(\ref{RKeqA}), 
\begin{equation}\label{RKeqD}
{\D R\over\D t} = - H_0 (1+z) \sqrt{Q}\, {\D R\over\D z}\quad .
\end{equation}
From the definition of $\Omega$ (Eq.~(\ref{HOL-g})) and matter
conservation (Eq.~(\ref{conserve-g})) we obtain 
\begin{equation}\label{RKeqE}
4\pi G\rho = {3\over 2}\,H_0^2\Omega_0 (1+z)^3\quad.
\end{equation}
If we now insert Eqs.~(\ref{RKeqB}), (\ref{RKeqC}), (\ref{RKeqD}) and
(\ref{RKeqE}) into Eq.~(\ref{zds-g}), sort the terms appropriately and
cancel $H_0^2$, which appears in all terms, we obtain 
\begin{equation}\label{RKeqF}
Q\, l'' + \left({2Q\over 1+z} + {1\over 2}\, Q'\right)\,l'
      + {3\over 2}\,\eta\,\Omega_0 (1+z)\, l = 0 \quad ,
\end{equation}
where a prime denotes a derivative with respect to redshift and from
Eq.~(\ref{q-g}) follows 
\begin{equation}
Q'(z) = 3\Omega_0(1+z)^2 - 2(\Omega_0+\lambda_0-1)(1+z)\quad .
\end{equation}
From the definition of the angular size distance (Eq.~(\ref{d-g})) it is
obvious that it follows the same differential equation as $l$: 
\begin{equation}\label{RKeqG}
\fbox{$\displaystyle
Q\, D'' + \left({2Q\over 1+z} + {1\over 2}\, Q'\right)\,D'
      + {3\over 2}\,\eta\,\Omega_0 (1+z)\, D = 0
$}
\end{equation}
with special boundary conditions at the redshift $z_x$ where the two
considered light rays intersect. The first boundary condition is
trivially 
\begin{equation}\label{boundary1-g}
D = 0 \quad {\rm for}\quad z = z_{x}\quad,
\end{equation}
and the second boundary condition follows from the Euclidean
approximation for small distances, i.e. 
\begin{equation}\label{boundary2-g}
\left. \,{\D D\over\D t}\right|_{z=z_x} = c \sign(t_x-t_y),
\end{equation}
hence
\begin{equation}\label{RKeqH}
D' = \frac{c}{H_{0}}\frac{1}{(1+z_x)\sqrt{Q(z_x)}}
     \sign(t_y-t_x) \quad {\rm for}\quad z=z_{x},
\end{equation}
where the sign has been chosen such that $D$ is always $>0$ locally. We
denote these special solutions of Eq.~(\ref{RKeqG}) with $D_x(z)$, and,
following the definition (Eq.~(\ref{d-g})), the angular size distance of
an object at redshift $z_y$ is then given as 
\begin{equation}\label{RKeqI}
D_{xy} = D_x(z_y) \quad. 
\end{equation}

\begin{figure}[t]
\epsffile{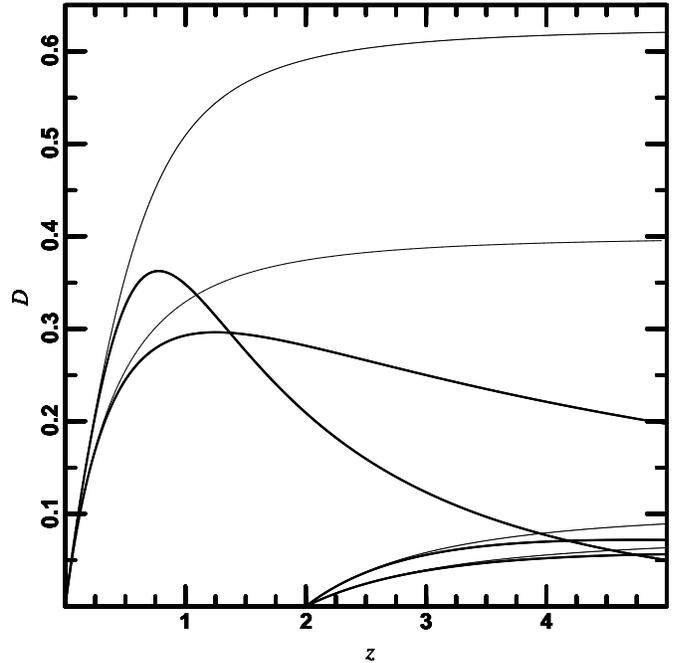}
\caption[]{The angular size distance from the observer ($z_{1}=0$) and
from $z_{1}=2$ (lower right) as a function of the redshift $z_{2}$ for
different cosmological models. Thin curves are for $\eta=0$, thick for
$\eta=1$.  The upper curves near $z=0$ ($z=2$ at lower right) are for
$\lambda_{0}=2$, the lower for $\lambda_{0}=0$.  $\Omega_{0}=1$ for all
curves.  The angular size distance~$D$ is given in units of $c/H_{0}$} 
\label{angsiz-f}
\end{figure}

Figure~\ref{angsiz-f} shows the influence of $z$, $\eta$ and $\lambda$
on the angular size distance, calculated using Eq.~(\ref{RKeqG}) with
our numerical implementation. 

For completeness we note that after the original derivation by Kayser
(\cite{RKayser85a}) an equivalent equation was derived by Linder
(\cite{ELinder88a}) which, however, is difficult to implement due to the
cumbersome notation. 

Special mention must be made of the so-called bounce models, which
expand from a finite $R$ after having contracted from $R=\infty.$ (See,
e.g., Feige (\cite{BFeige92a}).)  A glance at Eq.~(\ref{lemaitre-g})
shows that in these cosmological models there must be {\em four\/}
distances for an (ordered) pair of redshifts. If we denote the distances
by $D_{12}$, $D_{14}$, $D_{34}$ and $D_{32}$, where 1(2) und 3(4) refer
to $z_{1}$($z_{2}$) during the expanding (contracting) phase, then
symmetry considerations dictate that $D_{12}=D_{34}$ and $D_{14}=D_{32}$
{\em as long as the dependence of $\eta$ on $z$ is the same during both
phases}. In this case, there are {\em two\/} independent distances per
(ordered) pair of redshifts.  If this is not the case, the degeneracy is
no longer present and there are {\em four\/} independent distances per
(ordered) pair of redshifts.

\section{Numerics and practical considerations}\label{numerics}

For the actual numerical integration of the differential equation, we
have found the Bulirsch-Stoer method to be both faster and more exact
than other methods such as Runge-Kutta.  However, the conventional
method of rational function extrapolation is rather unstable in this
particular case; fortunately, using polynomial extrapolation solves the
problem.  Although programming the integration is rather straightforward
in theory, in numerical practice considerable effort is needed to
determine combinations of free parameters which work for all cases.  We
have tested the finished programme intensively and extensively, for
example by comparing the results of calculations for $\eta=1$ (the value
of $\eta$ plays no special role in the integration of the differential
equation) with those in Refsdal et al.~(\cite{SRefsdalSdL67a}) or given
by the method of elliptical integrals as outlined in Feige
(\cite{BFeige92a}) and have used it in Kayser (\cite{RKayser95a}),
Helbig (\cite{PHelbig96a}) and Helbig \& Kayser
(\cite{PHelbigRKayser96a}).  For a general discussion of various methods
of integrating second-order differential equations, see Press et~al.\
(\cite{WPressTVF92a}).  Those interested in technical details can read
the comments in our source code and the accompanying user's guide.

Since $H_{0}$, in contrast to the other cosmological parameters, merely
inversely scales the angular size distance, our routine actually
calculates the angular size distance in units of $c/H_{0}$.  This {\em
dimensionless\/} quantity must be multiplied by $c/H_{0}$ (in whatever
units are convenient) in order to obtain the actual distance.   Other
than reducing numerical overhead, this allows all distances to be
calculated modulo $c/H_{0}$, which is convenient for expressing
quantities in an $H_{0}$-independent manner.  In practice, $H_{0}$
cancels out of many calculations anyway. 

Apart from auxiliary routines which the user does not have to be
concerned with, our implementation consists of four {\tt FORTRAN77}
subroutines.  The first, {\tt INICOS}, calculates $z$-independent
quantities used by the other routines, some of which are returned to the
calling programme.  {\tt ANGSIZ} calculates the angular size distance.
Normally, $\eta$ is used as a $z$-independent cosmological parameter, on
an equal footing with $\lambda_{0}$ and $\Omega_{0}$.  If desired,
however, the user can let {\tt INICOS} know that a variable (that is,
$z$-dependent) $\eta$ is to be used; this is given by the function {\tt
VARETA}.  We supply an example; the user can modify this to suit her
needs.  In particular, many different dependencies of $\eta$ on $z$ can
be included, and a decision made in the calling programme about which
one to use.  This feature is also included in our example.  {\tt ANGSIZ}
returns only the distance $D_{12}$; if one is interested in the other
distances in the bounce models, our subroutine {\tt BNGSIZ} returns all
of these (though internally calculating only the independent distances,
of course, depending on the dependence of $\eta$ on $z$). 

Due to the fact that not everyone has a Fortran90 compiler at his
disposal, we have coded the routines in {\tt FORTRAN77}.  Only standard
{\tt FORTRAN77} features are used, and thus the routines should be able
to be used on all platforms which support {\tt FORTRAN77}.  Since
standard {\tt FORTRAN77} is a subset of Fortran90, the routines can be
used without change in Fortran90 as well. 

With the exception of $D^{\rm c}$, all distance measures can be easily
transformed into one another.  Thus, it suffices to calculate the
angular size distance for a given case.\footnote{The proper distance,
which is $\eta$-independent, can be calculated from the angular size
distance assuming $\eta=1$, by making use of the simple relation between
proper distance and angular size distance in this case.  The result
holds of course for all values of $\eta$.} 

When discussing the distance between two objects other than the
observer, rather than between the observer and one object, in many cases
one of two simplifying assumptions can be made: 
\begin{description}
\item[$D(\Delta z) \ll D(\beta)$] In this case, the {\em proper distance
$D^{\rm P}$ at the time of emission\/} between the two objects is $\beta
D_{0x}\approx \beta D_{0y}$, where $\beta \ll 1$ is the angle in radians
between the two objects on the sky. 
\item[$D(\beta) \ll D(\Delta z)$] In this case, the {\em angular size
distance\/} between the two objects is $D_{xy}$. 
\end{description}
$D(\Delta z)$ ($D(\beta$)) refers to the distance due to $\Delta z$
($\beta$) when setting $\beta$ ($\Delta z$) equal to zero. In the first
case, where the two objects are practically at the same redshift, one
uses the angular size distance to this redshift to transform the
observed difference in angular position on the sky into the {\em proper
distance\/} between the two objects at the time of emission.  This
follows directly from the definition of the angular size distance. 
Since the distance between the objects is much less than the distance
from the observer to the objects, the differently defined distances
between the objects are for practical purposes degenerate.  A practical
example of this case would be the distance between individual galaxies
in a galaxy cluster at large redshift. Naturally, one should use {\em
one\/} redshift, say, of the cluster centre; the individual redshifts
will in most cases be overlaid with the doppler redshift due to the
velocity dispersion of the cluster, so the difference in {\em
cosmological\/} redshifts is negligible.  (Of course, the {\em
present\/} distance would be a factor of $(1 + z)$ larger, due to the
expansion of the universe, were the objects comoving and not, as in a
galaxy cluster, bound.)  In the second case, which is typical of
gravitational lensing, the angles on the sky between, for example,
source and lens, are small enough to be neglected, so that the angular
size distance between the objects is determined by the difference in
redshift.  If neither of these assumptions can be made, any sort of
distance between the two objects is probably of no practical interest.
(Of course, there is the trivial case where the redshifts are all $\ll
1$ in which case one can simply use $\alpha$, $\delta$ and $cz/H_{0}$ as
normal spherical coordinates.)

\section{Summary}\label{summary}

After discussing cosmological distances with an emphasis on practical
distance measures for general use in cosmology and extragalactic
astronomy, we have obtained a new differential equation, which gives the
angular size distance for a class of `on average' Friedmann-Lema\^\i tre
cosmological models, that is, models described not only by $\lambda_{0}$
and $\Omega_{0}$ but also by $\eta(z)$, which describes the clumpiness
of the distribution of matter.  We have also developed a practical
numerical method of solving this equation, which we have made publicly
available.  Since the equation is valid for {\em all\/} cases, this
offers for the first time an efficient means of calculating distances in
a large class of cosmological models. 

The numerical implementation (in {\tt FORTRAN77}), user's guide and a
copy of the latest version of this paper can be obtained from either of 
the following URLs:
\begin{quote}
\scriptsize\tt http://www.hs.uni-hamburg.de/english/persons/helbig/ \\
Research/Publications/Info/angsiz.html
\end{quote}
\begin{quote}
\scriptsize\tt \tt ftp://ftp.uni-hamburg.de/pub/unihh/astro/angsiz.tar.gz
\end{quote}


\begin{acknowledgements}
It is a pleasure to thank O.~Czoske, S.~Refsdal and A.~Smette for
helpful discussions and comments on the manuscript. 
\end{acknowledgements}


\appendix

\section{Symmetry: The relation between $D_{xy}$ and $D_{yx}$}
\label{symmetry}

The proof in this appendix follows closely the proof presented in Kayser
(\cite{RKayser85a}).  For completeness we note that after the original
derivation by Kayser (\cite{RKayser85a}) an equivalent equation was
derived by Linder (\cite{ELinder88a}). We rewrite the differential
equation, Eq.~(\ref{RKeqG}), for the angular size distance in the normal
form: 
\begin{equation}\label{RKAeqA}
a_{2}D''(z) + a_{1}(z)\,D'(z) + a_{0}(z)\,D(z) = 0
\end{equation}
with the coefficient functions
\begin{equation}\label{RKAeqB}
a_{2}(z) = Q(z)
\end{equation}
\begin{equation}\label{RKAeqBB}
a_1(z) = {2Q(z)\over 1+z} + {1\over 2}\,Q'(z)
\end{equation}
\begin{equation}\label{RKAeqBBB}
a_{0}(z) = {3\over 2}\,\eta\,\Omega_0(1+z)\quad.
\end{equation}
Now let $D^{(1)}$ and $D^{(2)}$ be two solutions of Eq.~(\ref{RKAeqA})
which build a fundamental system, i.e.~the Wronskian for these two
solutions does not vanish: 
\begin{equation}
W(z) = \left|
    \begin{array}{cc}
     D^{(1)} & D^{(2)} \\
     {\D D^{(1)}\over \D z} & 
     {\D D^{(2)}\over \D z}
    \end{array}
    \right| \neq 0 \quad \forall z \quad.
\end{equation}
Every solution $D_i$ of Eq.~(\ref{RKAeqA}) can then be written as a
linear combination of $D^{(1)}$ and $D^{(2)}$: 
\begin{equation}\label{RKAeqD3}
D_i = \alpha_{i}\,D^{(1)} + \beta_{i}\,D^{(2)};
\quad{\rm with}\,\,\alpha_{i},\,\beta_{i}={\rm const}\quad.
\end{equation}
The angular size distances are special solutions $D_x$ of
Eq.~(\ref{RKAeqA}) fulfilling the following boundary conditions: 
\begin{equation}
D_x = 0 \quad {\rm for}\quad z=z_x
\end{equation}
and
\begin{equation}
{\D D_x\over\D z} = b(z_x)\quad{\rm for}\quad z=z_x
\end{equation}
with
\begin{equation}\label{RKAeqG}
b(z_x) = {c\over H_0}\,\left((1+z_x)\sqrt{Q(z_x)}\right)^{-1}
          \sign(t_y-t_x) \quad,
\end{equation}
compare Eq.~(\ref{RKeqH}). From Eq.~(\ref{RKAeqD3}) we obtain 
\begin{equation}
0 = \alpha_i\,D^{(1)}(z_x) + \beta_i\,D^{(2)}(z_x)
\end{equation}
and
\begin{equation}
b(z_x) = \alpha_i\,\left.{\D D^{(1)}\over\D z}\right|_{z=z_x}
      \, +\, \beta_i\, \left.{\D D^{(2)}\over\D z}\right|_{z=z_x}
\quad.
\end{equation}
These equations can easily be solved for $\alpha_i$ and $\beta_i$:
\begin{equation}
\alpha_i=\beta_i\,{D^{(2)}(z_x)\over D^{(1)}(z_x)}
\end{equation}
\begin{equation}
\beta_i={b(z_x)\,D^{(1)}(z_x)\over W(z_x)}
\end{equation}
and inserting $\alpha_i$ and $\beta_i$ back  into Eq.~(\ref{RKAeqD3}) we
obtain for the special solutions $D_x$: 
\begin{eqnarray}
D_x(z) &=& {b(z_x)\over W(z_x)}\,
  \left( D^{(1)}(z_x)\,D^{(2)}(z)\right.\nonumber\\ 
    & &\left. -\,  D^{(1)}(z)\,D^{(2)}(z_x)\right)
\quad.
\end{eqnarray}
If we now consider a second special solution $D_y$ we find the relation 
\begin{equation}
{D_x(z_y)\over D_y(z_x)} = -\,{b(z_x)\over b(z_y)}\,
    {W(z_y)\over W(z_x)}\quad.
\end{equation}
The Wronskians can be calculated using Liouville's formula:
\begin{equation}
W(z) = W(z_0)\,\exp\int\limits_z^{z_0} a_{2}(z)\,\D z
\quad,
\end{equation}
where $z_0$ is arbitrary. Thus
\begin{equation}
{D_x(z_y)\over D_y(z_x)} = -\,{b(z_x)\over b(z_y)}\,
\exp\int\limits_{z_y}^{z_x} \frac{a_1(z)}{a_{2}(z)}\,\D z
\quad
\end{equation}
and after inserting $a_{0}$, $a_1$ and $a_{2}$ from Eqs.~(\ref{RKAeqB}),
(\ref{RKAeqBB}) and (\ref{RKAeqBBB}) as well as $b(z_x)$ and $b(z_y)$
from Eq.~(\ref{RKAeqG}) and integration we finally obtain for the
angular size distances (cf.~Eq.~(\ref{RKeqI})) the relation 
\begin{equation}
{D_{xy}\over D_{yx}} = {1+z_x\over 1+z_y}\quad. 
\end{equation}


\section{Special cases}\label{special}

For certain special cases the differential equation can be simplified
and sometimes analytically solved.

\subsection{$\Omega_{0} = 0$}\label{omeganull}

A glance at Eq.~(\ref{RKeqG}) shows that for $\Omega_{0}=0$ the third
term on the left hand side of Eq.~(\ref{RKeqG}) vanishes; one thus has a
first order differential equation for $D'$. (Of course $\eta$ has no
meaning for $\Omega_{0}=0$.)  Due to the fact that a vanishing
$\Omega_{0}$ also simplifies $Q(z)$, it is possible to calculate the
angular size distance analytically.  Since in this case the angular size
distance is determined exclusively by global effects, one can use an
approach based on global geometry.\footnote{See the discussion in
Sect.~\ref{eta1}.}  Depending on the value of $\lambda_{0}$, one can use the 
following expression to calculate 
$\chi_{xy} = \chi(y) - \chi(x)$ (Feige \cite{BFeige92a}) 
\begin{equation}\label{omeganull-g}
\chi(z) = 
\left\{ \begin{array}{rcrr}
\arccosh(\psi) & & \rm for &     \lambda_{0} < 0 \\& & \\
\ln(1+z)       & & \rm for &     \lambda_{0} = 0 \\& & \\
\arcsinh(\psi) & & \rm for & 0 < \lambda_{0} < 1 \\& & \\
z              & & \rm for &     \lambda_{0} = 1 \\& & \\
\arcsin(\psi)  & & \rm for &     \lambda_{0} > 1 
\end{array}\right.,
\end{equation} 
where $\psi:=(1+z)\sqrt{\frac{|1-\lambda_{0}|}{|\lambda_{0}|}}$.  The
relationship between $\chi$ and the angular size distance~$D$ is 
\begin{equation}
D_{xy} = \frac{R_{0}}{(1 + z_{y})}\left\{ \begin{array}{rcrrrr}
\sinh\chi & & \rm for & k & = & -1 \\
        \chi & & \rm for & k & = &  0 \\
\sin \chi & & \rm for & k & = & +1 
\end{array}\right.,
\end{equation}
as discussed below in Sect~\ref{eta1}.

\subsection{$\eta = 0$}

In the case $\eta=0$ the third term on the left hand side of
Eq.~(\ref{RKeqG}) vanishes; one thus has a first order differential
equation for $D'$.  Assuming $D'\neq0$,  Eq.~(\ref{RKeqG}) can be
written as 
\begin{equation}
\frac{D''}{D'}=-\frac{2}{1+z}-\frac{1}{2}\frac{Q'(z)}{Q(z)}\quad.
\end{equation}
This equation can be solved in two steps.  For $D'$ we obtain
\begin{equation}
D'={\frac {c_1}{\sqrt{Q(z)}\left (1+z\right )^2}}
\end{equation} 
and consequently for $D$
\begin{equation}
D = \int{\frac {c_1}{\sqrt{Q(z)}\left (1+z\right )^2}}+c_2\quad.
\end{equation}  
The constants $c_1, c_2$ are determined by the appropriate boundary
conditions (Eqs.~(\ref{boundary1-g}) and (\ref{boundary2-g})). We then
find the solution (see also SEF for an equivalent discussion with
$\lambda_{0}=0$) 
\begin{equation}\label{domega-g}
D_{xy} = \frac{\rm c}{{\rm H}_0}(1 + z_{x})(\omega(z_{y})-\omega(z_{x})), 
\end{equation}
where 
\begin{equation}\label{omega-g}
\omega(z) =              \int\limits_{0}^{z} \frac{{\rm d}z'}
                      {(1 + z')^{2}\sqrt{(1+z')^{2}(\Omega_{0}z'+1-
\lambda_{0})+\lambda_{0}}},
\end{equation}
or, perhaps more convenient, 
\begin{equation}
D_{xy}=\frac{c}{{H}_0}\left(1+z_x\right)\int_{z_x}^{z_y}\!
       \frac{{\rm d}z}{(1+z)^{2}\sqrt{Q(z)}}\quad.
\end{equation}

For $\lambda_{0}=0$ there is an analytic solution (see
Sect.~\ref{lambdanull}).

\subsection{$\eta = 1$}\label{eta1}

The case $\eta=1$ has all matter distributed homogeneously. Due to
homogeneity, the matter {\em locally\/} affecting the propagation of
light is known when the {\em global\/} geometry is known, so that the
`classical' approach of relating global geometry to observable relations
is a better approach than using (the simplified form of)
Eq.~(\ref{RKeqG}).  This approach offers an analytic solution. Here, we
simply sketch the most important points; the interested reader can refer
to Feige (\cite{BFeige92a}) for a good description of this method. 

The angular size distance in this case is 
\begin{equation}\label{ang-g}
D_{xy} = R_{y}\sigma_{xy} = \frac{R_{0}\sigma_{xy}}{(1 + z_{y})},
\end{equation} 
where $\sigma$ is the radial coordinate in the Robertson-Walker metric
(cf.~Eq.~(\ref{rwm-g})) and thus
\begin{equation}
D_{yx} = \frac{R_{0}\sigma_{xy}}{1 + z_{x}} = 
D_{xy}\left(\frac{1 + z_{y}}{1 + z_{x}}\right),
\end{equation}
since this angle is inversely proportional to R for constant $\sigma$
and physical size.  (The value of $R$ at the time the light rays
defining the angle intersect is important.) 

Since $\sigma$ is given by 
\begin{equation}\label{sigma-g}
\sigma = F(\chi) = \left\{ \begin{array}{rcrrrr}
\sinh\chi & & \rm for & k & = & -1 \\
        \chi & & \rm for & k & = &  0 \\
\sin \chi & & \rm for & k & = & +1 
\end{array}\right.,
\end{equation} 
an expression for $\chi(z)$ is sufficient for calculating the angular
size distance~$D$ (and of course the luminosity distance~$D^{\rm L}$
(via Eq.~(\ref{lumang-g})) and the `coordinate distance'~$\sigma$ (via
Eq.~(\ref{sigma-g})).  In general, $\sigma_{xy} \neq \sigma_{y} -
\sigma_{x}$; however, $\chi_{xy} = \chi_{y} - \chi_{x}$, so that 
\begin{equation}
\sigma_{xy} = F (\chi_{xy})
\end{equation}
where $F$ is given by Eq.~(\ref{sigma-g}).  Using Eq.~(\ref{dp-g}) one
can calculate 
\begin{equation}
\chi_{xy} = \frac{D^{P}}{R_{0}} = 
\frac{c}{H_{0}R_{0}}\int\limits_{z_{x}}^{z_{y}}\frac{{\rm d}z}{\sqrt{Q(z)}}.
\end{equation}
In the general case, Eq.~(\ref{dp-g}) can be solved by elliptic
integrals, as explained in Feige (\cite{BFeige92a}).  For the cases
$\lambda_{0}=0$ and $\Omega_{0}=0$ the formulae using elliptic integrals
break down; in these cases, easier analytic formulae, which fortunately
exist, can be used.  The case $\Omega_{0}=0$ has been discussed above. 
The case $\lambda_{0}=0$ will be discussed below. {\em Again, we stress
that the differential equation derived in Sect.~\ref{rainer} is
completely general and can be used in {\em all} cases}.

\subsection{$\lambda_{0} = 0$}\label{lambdanull}

For $\lambda_{0} = 0$, there is in general no simpler solution. This
case has been discussed by Dyer and Roeder for $\eta=0$
(\cite{CDyerRRoeder72a}) and for general $\eta$ values
(\cite{CDyerRRoeder73a}).  They point out the interesting result that
the maximum in the angular size distance from $z_{1}=0$ to $z_{2}$
increases monotonically from $1.25$ to $\infty$ as $\eta$ decreases from
$1$ to $0$.  See also the discussion (with a differing notation!) in
Sect.~4.5.3 in SEF.  However, some solutions exist for special values of
$\Omega_{0}$ and $\eta$.  The case $\Omega_{0}=0$ has been discussed in
Sect.~\ref{omeganull} above; the value of $\eta$ is of course irrelevant
in this case.   With the exception of Sect~\ref{mattig} below, in the
following we simply quote results from SEF in our notation.

\subsubsection{$\lambda_{0} = 0$ and $\eta = 0$}

As discussed above, for $\eta=0$ Eq.~(\ref{RKeqG}) is effectively a
first order equation for $D'$.  For $\lambda_{0}=0$ $Q(z)$ is
sufficiently simplified to allow an analytic solution.  Recalling
Eq.~(\ref{domega-g}), 
\begin{displaymath}
D_{xy} = \frac{\rm c}{{\rm H}_0}(1 + z_{x})(\omega(z_{y})-\omega(z_{x})), 
\end{displaymath}
Eq.~(\ref{omega-g}) simplifies to 
\begin{equation}
\omega(z) = \int\limits_{0}^{z} \frac{{\rm d}z'}
                      {(1 + z')^{3}\sqrt{\Omega_{0}z'+1}},
\end{equation}
which has the solution:
\begin{equation}
\omega(z) =
\left\{\begin{array}{lr}
\frac{3\Omega_{0}^{2}}{4(\Omega_{0}-1)^\frac{5}{2}} 
\arctan(\psi) + \frac{3}{4(\Omega_{0}-1)^{2}} \quad \times & \\ 
\left(
\frac{\left(\Omega_{0}z+\frac{5\Omega_{0}}{3}-\frac{2}{3}\right)
\sqrt{\Omega_{0}z+1}}{(1+z)^{2}}-\frac{5\Omega_{0}}{3}+\frac{2}{3}\right)
                                             &  (\rm I)\\
\quad & \\
\frac{2}{5}\left(1 - (\Omega_{0}z+ 1 )^{-\frac{5}{2}}\right)
                                             &  (\rm II) \\
\quad & \\
\frac{3\Omega_{0}^{2}}{4(1-\Omega_{0})^\frac{5}{2}} 
\arctanh(\psi) + \frac{3}{4(1-\Omega_{0})^{2}} \quad \times & \\ 
\left(
\frac{\left(\Omega_{0}z+\frac{5\Omega_{0}}{3}-\frac{2}{3}\right)
\sqrt{\Omega_{0}z+1}}{(1+z)^{2}}-\frac{5\Omega_{0}}{3}+\frac{2}{3}\right)
                                             &  (\rm III)
\end{array}\right.
\end{equation}
with
\begin{equation}
\psi =
\left\{ \begin{array}{lr}
 \left(\frac{\sqrt{\Omega_{0}-1}(1 + \sqrt{\Omega_{0}z+1})}
{\Omega_{0}-1+\sqrt{\Omega_{0}z+1}}\right)
                                                &(\rm I) \\&  \\
\left(\frac{\sqrt{1-\Omega_{0}}(1 + \sqrt{\Omega_{0}z+1})}
{\Omega_{0}-1+\sqrt{\Omega_{0}z+1}}\right)
                                                & (\rm III)
\end{array}\right.
\end{equation}
and
\begin{equation}
\begin{array}{lr}
\mbox{case I:}                      &      \Omega_{0} > 1 \\
\mbox{case II:}                     &      \Omega_{0} = 1 \\
\mbox{case III:}                    &  0 < \Omega_{0} < 1 
\end{array}.
\end{equation}
Note that in SEF, the text at the top of page~137 is unclear---the
expression in parentheses in the denominator of the first term
($\Omega-1$) for the $\Omega>1$ case has to be replaced with
($1-\Omega$) as well for $\Omega<1$.  Note also that
$\Omega\equiv\Omega_{0}$ and that after page 131 $\lambda_{0}=0$ is
always assumed.

\subsubsection{$\lambda_{0} = 0$ and $\Omega_{0} = 1$}

For $\Omega_{0}=1$ and $\lambda_{0}=0$ (the Einstein-de~Sitter model) we 
have the solution 
\begin{equation}
D_{xy}=\frac{c}{H_{0}}\frac{1}{2\beta}\left(\frac{(1+z_{y})^{\beta-\frac{5}{4}}}
{(1+z_{x})^{\beta+\frac{1}{4}}}-\frac{(1+z_{x})^{\beta-\frac{1}{4}}}
{(1+z_{y})^{\beta+\frac{5}{4}}}\right),
\end{equation}
where 
\begin{equation}
\beta:=\frac{1}{4}\sqrt{25 - 24\eta}.
\end{equation}

\subsubsection{$\lambda_{0} = 0$ and $\eta = 1$}\label{mattig}

For $\eta=1$ the special case of the expression for $\chi(z)$ for
$\lambda_{0}=0$ is (Feige~\cite{BFeige92a}) 
\begin{equation}
\chi(z) = 
-2\left\{ \begin{array}{lr}
\arcsin\left(\sqrt{\frac{\Omega_{0}-1}{\Omega_{0}(1+z)}}\right)
                                              &   (  \Omega_{0} > 1 ) \\
&  \\
\sqrt{\frac{1}{1+z}}
                                            &   (  \Omega_{0} = 1 ) \\
&  \\      
\arcsinh\left(\sqrt{\frac{1-\Omega_{0}}{\Omega_{0}(1+z)}}\right)
                                              & (0 < \Omega_{0} < 1 )
\end{array}\right.,
\end{equation}
where $\chi_{xy} = \chi(y) - \chi(x)$. 
(It is obvious that in the case $\lambda_{0}=\Omega_{0}=0$
Eq.~(\ref{omeganull-g}) should be used.)  From this, it is possible to
obtain a general expression for the angular size distance (see, e.g.,
SEF): 
\begin{equation}
D_{xy} = \frac{c}{H_{0}}\frac{2}{\Omega_{0}^{2}}(1+z_{x})\left(
R_{1}(z_{y})R_{2}(z_{x}) - R_{1}(z_{x})R_{2}(z_{y})\right),
\end{equation}
with
\begin{equation}
R_{1}(z) = \frac{\Omega_{0}z - \Omega_{0} + 2}{(1+z)^{2}}
\end{equation}
and
\begin{equation}
R_{2}(z) = \frac{\sqrt{\Omega_{0}z+1}}{(1+z)^{2}}.
\end{equation}
For $z_{x}=0$ and $z_{y}=z$ one gets for the angular size distance
\begin{eqnarray}
\label{msef-g}
D(z)& = & \frac{c}{H_{0}}
\frac{2}{\Omega_{0}^{2}(1+z)^{2}} \qquad\times\nonumber \\
 & &
\left(
\Omega_{0}z - (2 - \Omega_{0})\left(\sqrt{\Omega_{0}z+1}-1\right)\right).
\end{eqnarray}
valid for $\Omega_{0}>0$.  For $\Omega_{0}=0$ one obtains
\begin{equation}
\label{msef0-g}
D = \frac{c}{H_{0}}\frac{z\left(1 + \frac{z}{2}\right)}{\left(1 + z\right)^{2}}
\end{equation} 
(Multiplying Eq.~(\ref{msef-g}) or Eq.~(\ref{msef0-g}) with
$(1+z)/R_{0}$ results in the respective expression for $\sigma$ as a
function of redshift as first derived by Mattig (\cite{WMattig58a}). See
also Sandage (\cite{ASandage95a}), Sect.~1.6.3). In this case, the
volume element given by Eq.~(\ref{dvolume-g}) reduces to 
\begin{equation}
\D V = 16\pi R_{0}^{3}
\frac{(\Omega_{0}z -(2-\Omega_{0})(\sqrt{\Omega_{0}z+1}-1)^{2})}
{\Omega_{0}^{4}(1+z)^{3}\sqrt{\Omega_{0}z+1}}
\end{equation}
Of course, for the physical, as opposed to comoving, density, an {\em
additional\/} factor of $(1+z)^{3}$ must be added to the denominator. 

\subsubsection{$\lambda_{0} = 0$ and $\eta = \frac{2}{3}$}

For $\eta=\frac{2}{3}$ and $\lambda_{0}=0$ there is also an analytic
solution (see SEF): 
\begin{eqnarray}
D_{xy} & = &
\frac{c}{H_{0}}\frac{2}{3\Omega_{0}^{2}}(1+z_{x})\qquad\times\nonumber\\
 & &
\left(R_{1}(z_{x})R_{2}(z_{y}) - R_{2}(z_{x})R_{1}(z_{y})\right),
\end{eqnarray}
with
\begin{equation}
R_{1}(z) = \frac{1}{(1+z)^{2}}
\end{equation}
and
\begin{equation}
R_{2}(z) = \frac{\sqrt{\Omega_{0}z+1}(\Omega_{0}z+3\Omega_{0}-2)}{(1+z)^{2}}.
\end{equation}

\subsection{Other cases}

We can offer no proof that no other easier solutions, either reducing
Eq.~(\ref{RKeqG}) to a more easily (numerically) integrated form or even
to an analytic solution, exist.  This is left as an exercise to the
interested reader.  The authors are of course interested in such
solutions and are willing to verify them.  As far as we know,
Eq.~(\ref{RKeqG}) must be used except in the special cases mentioned in
this appendix.

\subsection{Light travel time}\label{time}

Feige (\cite{BFeige92a}) not only gives the distance but also the light
travel time by means of elliptic integrals.  As for the distance, and
for the same reasons, simple analytic formulae can and must be used for
the special cases $\Omega_{0}=0$ and $\lambda_{0}=0$.  For $k=0$, an
analytic expression for the light travel time exists, although the
elliptic integrals can also be used in this case.  For completeness, we
give these special cases here for the light travel time $t_{xy} = t_{x}-t_{y}$. 

For $\Omega_{0}=0$ we have:
\begin{equation}\label{tomeganull-g}
t(z) = \frac{1}{H_{0}\sqrt{|\lambda_{0}|}}
\left\{ \begin{array}{lr}
\arcsin(\psi)                         & \lambda_{0} < 0 \\&  \\
\frac{\sqrt{|\lambda_{0}|}}{(1+z)}   & \lambda_{0} = 0 \\&  \\
\arcsinh(\psi)                        & 0 < \lambda_{0} < 1 \\ & \\
-\sqrt{|\lambda_{0}|}\ln(1+z)         & \lambda_{0} = 1 \\&  \\
\arccosh(\psi)                        & \lambda_{0} > 1
\end{array}\right.,
\end{equation} 
where $\psi:=\frac{1}{(1+z)}\sqrt{\frac{|\lambda_{0}|}{|1-\lambda_{0}|}}$ 

For $\lambda_{0}=0$ we have:
\begin{equation}
t(z) = A \times
\left\{ \begin{array}{lr}
\frac{\sqrt{(\Omega_{0}z+1)(\Omega_{0}-1)}}{\Omega_{0}(1+z)} \quad - & \\
\arcsin\left(\sqrt{\frac{\Omega_{0}-1}{\Omega_{0}(1+z)}}\right)
                                                       & \Omega_{0} > 1  \\ 
\qquad\\
-\frac{\sqrt{\Omega_{0}-1}^{3}}{\Omega_{0}} 
\frac{2}{3}\left(\sqrt{\frac{1}{1+z}}\right)^{3} 
                                   & \Omega_{0} = 1                       \\
\qquad\\
\arcsinh\left(\sqrt{\frac{1-\Omega_{0}}{\Omega_{0}(1+z)}}\right) \quad - \\
\frac{\sqrt{(\Omega_{0}z+1)(1-\Omega_{0})}}{\Omega_{0}(1+z)}
                                      & 0 < \Omega_{0} < 1
\end{array}\right.,
\end{equation} 
where 
\begin{displaymath}
A = -\frac{\Omega_{0}}{H_{0}(\sqrt{|\Omega_{0}-1|})^{3}}
\end{displaymath}
(For $\Omega_{0}=0$ the appropriate case from Eq.~(\ref{tomeganull-g}) 
must be used.)  

For $k=0$ we have:
\begin{equation}
t(z) = \frac{2}{3H_{0}} \times
\left\{ \begin{array}{lr}
\frac{1}{\sqrt{\Omega_{0}-1}} 
\arcsin(\psi)
& \qquad     \Omega_{0} > 1 \\&  \\
\left(\sqrt{\frac{1}{1+z}}\right)^{3}       
   &    \qquad  \Omega_{0} = 1 \\ & \\
\frac{1}{\sqrt{1-\Omega_{0}}}
\arcsinh(\psi)
& \qquad 0 < \Omega_{0} < 1
\end{array}\right., 
\end{equation} 
where 
\begin{equation}
\psi = \left\{  \begin{array}{lr}
\sqrt{\frac{\Omega_{0}-1}{\Omega_{0}(1+z)^{3}}}
                   & \qquad\qquad \Omega_{0} > 1 \\&  \\
\sqrt{\frac{1-\Omega_{0}}{\Omega_{0}(1+z)^{3}}}
& \qquad\qquad 0 < \Omega_{0} < 1
\end{array}\right..
\end{equation} 
(For $\Omega_{0}=0$ the appropriate case from Eq.~(\ref{tomeganull-g}) 
must be used.)

\section{Volume element}

Sometimes the distance is only a means of calculating the volume element
at a given redshift.  In the static Euclidean case the volume element is
of course 
\begin{equation}
\D V = 4\pi r^{2}dr.
\end{equation}
In the cosmological case, the volume element is, with $r=R_{0}\sigma$,
\begin{equation}
\D V = 4\pi r^{2}\D D^{\rm P} = 
4\pi r^{2}\frac{c}{H_{0}}\frac{{\rm d}z}{\sqrt{Q(z)}}.
\end{equation}
$R_{0}\sigma_{y}$ is, for $\eta=1$, simply $D_{y0}=(1 + y)D_{0y}$; see
Sect~\ref{eta1}.  Thus, the distance $D_{y0}$ is all that is needed to
calculate the volume; this first can be calculated by Eq.~(\ref{RKeqG})
with $\eta=1$ (This applies even if one would calculate {\em
distances\/} with another value of $\eta$ since the volume element is a
quantity related to the global geometry of the universe---alternatively,
one can use elliptic integrals, as in Sect.~\ref{eta1} and Feige
(\cite{BFeige92a}).) If one has an expression for $\sigma(z)$, then,
since 
\begin{equation}
\D D^{\rm P} = R_{0}\frac{\D \sigma}{\sqrt{1-k\sigma^{2}}},
\end{equation}
which follows directly from Eq.~(\ref{rwm-g}), one can write
\begin{equation}
\label{dvolume-g}
dV(\sigma) = 4\pi R_{0}^{3}
\int\limits_{0}^{\sigma}\frac{\sigma'^{2}\D\sigma'}{\sqrt{1-k\sigma'^{2}}}.
\end{equation}
where $R_{0}$ is given by Eq.~(\ref{rnull-g}) for the present values:
\begin{equation}
R_{0} = \frac{c}{H_{0}}\frac{1}{\sqrt{|\Omega_{0} + \lambda_{0} -1|}}.
\end{equation}
and 
\begin{equation}
\sigma = \frac{D_{y0}(1+z_{y})}{R_{0}}
\end{equation}
Integration gives
\begin{equation}
\label{volume-g}
V(\sigma) = \left\{
\begin{array}{lr}
2\pi r^{3}\left(\frac{\sqrt{1+\sigma^{2}}}{\sigma^{2}} -
\frac{\arcsinh \sigma}{\sigma^{3}}  
\right)
& \qquad k = -1 \\
\frac{4}{3}\pi r^{3}
& \qquad k = 0 \\
2\pi r^{3}\left(\frac{\arcsin \sigma}{\sigma^{3}} - 
\frac{\sqrt{1-\sigma^{2}}}{\sigma^{2}}\right)
& \qquad k = +1
\end{array}
\right.
\end{equation}
Thus, for $k=+1$, the total volume of the universe is
$2\pi^{2}R_{0}^{3}$. (See, e.g., Sandage (\cite{ASandage95a}),
Sect.~1.6.1; Sandage's $d$ is our $D^{\rm P}$ and his $r$ is our
$\sigma$.) Since 
\begin{displaymath}
\D\chi=\frac{\D\sigma}{\sqrt{1-k\sigma^{2}}}
\end{displaymath}
Eq.~(\ref{volume-g}) can also be written as 
(cf.~Feige (\cite{BFeige92a}), Eq.~(116); Feige's $r$ is our $\sigma$)
\begin{equation}
\label{volume2-g}
V(\chi) = 2\pi R_{0}^{3}\left\{
\begin{array}{lr}
\sinh(\chi)\cosh(\chi) - \chi
& \qquad k = -1 \\
\frac{2}{3}\chi^{3}
& \qquad k = 0 \\
\chi - \sin(\chi)\cos(\chi)
& \qquad k = +1
\end{array}
\right.
\end{equation}
Of course, all this refers to volumes {\em now\/} at the distance
corresponding to $z=y$.  If the volume at another time is important, say
at the time of emission of the light we see now---for instance if one is
concerned with the space density of some comoving objects---then the volume
element must be divided by $(1 + z)^{3}$.




\begin{thebibliography}{}
\book{1986}{MBerry86a}{\firstauthor{M.~V.}{Berry}} %
{Cosmology and Gravitation}{Bristol}{Adam Hilger}{1986}
%
\book{1961}{HBondi61a}{\firstauthor{H.}{Bondi}} %
{Cosmology}{Cambridge}{Cambridge University Press}{1961}
%
\article{1980}{CColemanWW80a}{\firstauthor{G.~D.}{Coleman}, %
\otherauthor{C.-C.}{Wu}, \otherauthor{D.~W.}{Weedman}}%
{}%
{ApJS}{43}{3}{393}{1980}
%
\article{1966}{VDashevskiiVSlysh66a}{\firstauthor{V.~M.~}{Dashevskii}, %
\otherauthor{V.~J.}{Slysh}}%
{}{Sov.~Astr.}{9}{}{671}{1966}
%
\article{1965}{VDashevskiiYZeldovich65a}{\firstauthor{V.~M.~}{Dashevskii}, %
\otherauthor{Y.~B.}{Zeldovich}}%
{}{Sov.~Astr.}{8}{}{854}{1965}
%
\novolume{1972}{CDyerRRoeder72a}{\firstauthor{C.~C.}{Dyer}, %
\otherauthor{R.~C.}{Roeder}}%
{}
{ApJ}{174}{L115}{1972}
%
\novolume{1973}{CDyerRRoeder73a}{\firstauthor{C.~C.}{Dyer}, %
\otherauthor{R.~C.}{Roeder}}%
{}%
{ApJ}{180}{L31}{1973}
%
\article{1992}{BFeige92a}{\firstauthor{B.}{Feige}}%
{Elliptical integrals for cosmological constant cosmologies}%
{Astr. Nachr.}{313}{3}{139}{1992}
%
\article{1995}{AGoobarSPerlmutter95a}{\firstauthor{A.}{Goobar}, %
\otherauthor{S.}{Perlmutter}}{Feasibility of measuring the cosmological 
constant $\Lambda$ and mass density $\Omega$ using type Ia supernovae}%
{ApJ}{450}{1}{14}{1995}
%
\proceedings{1996}{PHelbig96a}{\firstauthor{P.}{Helbig}}%
{Predicted lens redshifts and magnitudes}%
{Astrophysical Applications of Gravitational Lensing (IAU Symposium 173)}%
{\firstauthor{C.~S.}{Kochanek}, \otherauthor{J.}{Hewitt}}%
{Dordrecht}{Kluwer}{1996}
%
\article{1996}{PHelbigRKayser96a}{\firstauthor{P.}{Helbig}, %
\otherauthor{R.}{Kayser}}%
{Cosmological parameters and the redshift distribution of gravitational %
lenses}%
{A\&A}{308}{2}{359}{1996}
%
\article{1983}{RKayserSRefsdal83a}{\firstauthor{R.}{Kayser}, %
\otherauthor{S.}{Refsdal}}%
{The difference in light travel time between gravitational lens images. 
inhomogeneous universes}%
{A\&A}{128}{?}{156}{1983}   
%
\doctor{1985}{RKayser85a}{\firstauthor{R.}{Kayser}}%
{Helligkeits\"{a}nderung durch den statistischen Gravitationslinseneffekt}%
{University of Hamburg}{1985}
%
\article{1995}{RKayser95a}{\firstauthor{R.}{Kayser}}%
{A cosmological test with compact radio sources}%
{A\&A}{294}{?}{L21}{1995}   
%
\novolume{1988}{ELinder88a}{\firstauthor{E.~V.}{Linder}}%
{Light propagation in generalized Friedmann universes}%
{A\&A}{206}{190}{1988}
%
\article{1993}{MLongair93a}{\firstauthor{M.}{Longair}}%
{Modern cosmology: a critical assessment}%
{QJRAS}{34}{2}{157}{1958}
%
\novolume{1958}{WMattig58a}{\firstauthor{W.}{Mattig}}%
{}%
{Astr.~Nachr.}{284}{109}{1958}
%
\article{1995}{SPerlmutter95a}{\firstauthor{S.}{Perlmutter}, %
\otherauthor{C.~R.}{Pennypacker}, \otherauthor{G.}{Goldhaber}, \etal{{%
\otherauthor{A.}{Goobar}, \otherauthor{R.~A.}{Muller}, %
\otherauthor{H.~J.~M.}{Newberg}, \otherauthor{J.}{Desai}, %
\otherauthor{A.~G.}{Kim}, \otherauthor{M.~Y.}{Kim}, %
\otherauthor{I,~A.}{Small}, \otherauthor{B.~J.}{Boyle}, %
\otherauthor{C.~S.}{Crawford}, \otherauthor{R.~G.}{McMahon} %
\otherauthor{P.~S.}{Bunchark}, \otherauthor{D.}{Carter}, %
\otherauthor{M.~J.}{Irwin}, \otherauthor{R.~J.}{Terlevich} %
\otherauthor{R.~S.}{Ellis}, \otherauthor{K.}{Glanzebrook}, %
\otherauthor{W.~J.}{Couch}, \otherauthor{J.~R.}{Mould} %
\otherauthor{T.A.}{Small}, \otherauthor{R.~G.}{Abraham}}}}%
{A supernova at $z=0.458$ and implications for measuring the %
cosmological deceleration}%
{ApJ}{440}{2}{L41}{1995}
%
\book{1992}{WPressTVF92a}{\firstauthor{W.~H.}{Press}, %
\otherauthor{S.~A.}{Teukolsky}, \otherauthor{W.~T.}{Vetterling}, %
\otherauthor{B.~P.}{Flannery}}%
{Numerical Recipes in FORTRAN}{Cambridge}{Cambridge University Press}{1992}
%
\novolume{1967}{SRefsdalSdL67a}{\firstauthor{S.}{Refsdal}, %
\otherauthor{R.}{Stabell}, \otherauthor{F.~G.}{de~Lange}}%
{Numerical calculations on relativistic cosmological models}%
{Mem.~R.~Astron.~Soc.}{71}{143}{1967}
%
\proceedings{1995}{ASandage95a}{\firstauthor{A.}{Sandage}}%
{Practical Cosmology: Inventing the Past}%
{The Deep Universe}%
{\firstauthor{B.}{Binggeli}, \otherauthor{R.}{Buser}}%
{Berlin}{Springer}{1995}
%
\book{1992}{PSchneiderEF92a}{\firstauthor{P.}{Schneider}, %
\otherauthor{J.}{Ehlers}, \otherauthor{E.~E.}{Falco}}%
{Gravitational Lenses}{Heidelberg}{Springer-Verlag}{1992}
%
\article{1976}{SWeinberg76a}{\firstauthor{S.}{Weinberg}} %
{Apparent luminosities in a locally inhomogeneous universe}%
{ApJ}{208}{1}{L1}{1976}
%
\article{1964}{YZeldovich64a}{\firstauthor{Y.~B.~}{Zeldovich}} %
{}{Sov.~Astr.}{8}{}{13}{1964}
%
\end{thebibliography}
\end{document}